\documentclass[conference]{IEEEtran}
\ifCLASSINFOpdf
\else
\fi
\usepackage{amsthm}
\usepackage{makeidx}
\usepackage{examples}
\usepackage[pdftex]{graphicx}
\usepackage{amssymb}
\usepackage{graphicx}
\usepackage{algorithm}
\usepackage{epstopdf}
\usepackage{color}
\usepackage{url}
\graphicspath{{../eps/}{../ps/}}
\usepackage{psfrag}
\usepackage{algorithm}
\usepackage{algorithmic}
\usepackage{amsmath}
\usepackage{hyperref}

\newtheorem{theorem}{Theorem}

\newtheorem{remark}[theorem]{Remark}

\newtheorem{definition}[theorem]{Definition}

\begin{document}
%
\title{DNACloud: A Tool for Storing Big Data on DNA}

\author{
\IEEEauthorblockN{Shalin Shah, Dixita Limbachiya and Manish K. Gupta}
\IEEEauthorblockA{Laboratory of Natural Information Processing\\
Dhirubhai Ambani Institute of Information and Communication Technology\\
Gandhinagar, Gujarat, 382007 India\\
Email: shalinshah1993@gmail.com, dlimbachiya@acm.org, mankg@computer.org}
}


%


\maketitle

\begin{abstract}
The term Big Data is usually used to describe huge amount of data that is generated by humans from digital media such as cameras, internet, phones, sensors etc. By building advanced analytics on the top of big data, one can predict many things about the user such as behavior, interest  etc. However before one can use the data, one has to address many issues for big data storage. Two main issues are the need of large storage devices and the cost associated with it.  Synthetic DNA storage seems to be an appropriate solution to address these issues of the big data. Recently in $2013$, Goldman and his collegues from European Bioinformatics Institute demonstrated the use of the DNA as storage medium with capacity of storing $1$ peta byte of information on one gram of DNA and retrived the data successfully with low error rate \cite{goldman2013towards}. This significant step shows a promise for synthetic DNA storage as a useful technology for the future data storage. Motivated by this, we have developed a software called DNACloud which makes it easy to store the data on the DNA. In this work, we present detailed description of the software.
\end{abstract}

%
\IEEEpeerreviewmaketitle

\section{Introduction}
Storage has been a fundamental requirement for the Humans. In the modern era of computing and communication, huge amount of data is being generated and there is a pressing need for dense storage medium which is cost effective. 
Table \ref{data}  shows the typical amount of the data generated and the kind of storage device it will require to store such a data.  It is predicted that by $2015$, the amount of data generated by NSA (National Security Agency) will be so large that it may need $1000$ billion tera bytes of hard disk space worth $\$1,000$ trillion \cite{CompareByte}. At present, the world is producing 1 exabytes of data per day and soon devices, machines and sensors of  Internet of Things (IoT) will generate data in the order of bronobytes, where $1$ bronobyte is $10^{27}$ bytes, \cite{reviewpaper} for which a dense storage 
medium is needed. From the past 30 years, the blue print of life viz. DNA has been used as storage medium. 
Unlike existing storage device, DNA requires no maintenance and can be stored without electricity in cold and dark place. One of the venture to use the DNA as artistic material and convert the graphic image to the language of genetic code was initiated by Joe Davis in the work Microvenus \cite{davis1996microvenus}.  In $1999$, Synthetic gene that was created by Kac by translating a sentence from the biblical book of Genesis into Morse Code, and converting the Morse code into DNA base pairs according to a conversion principle \cite{Genesis}. In the $20^{th}$ century, many researchers have translated English text, mathematical equations \cite{yachie2007alignment}, latin text \cite{portney2008length} and simple musical notations \cite{ailenberg2009improved} to DNA using different DNA coding principles \cite{wong2003organic} \cite{arita2004secret} \cite{skinner2007biocompatible}. All the above mentioned efforts were successful on a small scale giving birth to the idea of data storage on DNA. But the most prolific work was done in $2012$ by Church, et al. \cite{church2012next} of Harvard University. They encoded successfully entire book of \itshape Regenisis: How Synthetic Biology Will Reinvent Nature and Ourselves \normalfont\cite{church2012regenesis}  including $53,426$ words, $11$ JPG images and a JavaScript program into DNA using $1$ bit per base encoding. The main draw back of their method was that it had high error rate \cite{church2012next}. In the subsequent year, in $2013,$ this limitation was overcomed by the Goldman and his group. They implemented  a modified approach that includes error correction and scaled DNA based data storage \cite{goldman2013towards}. Based on this method of DNA data storage \cite{goldman2013towards}, in this work we present the software called DNACloud which converts the data file to DNA sequences and vice versa. The reader is referred to excellent short reviews of synthetic DNA storage \cite{pmid23514938,Greengard:2013:NAI:2492007.2492013} to get an overview of this new area.

This paper is organized as follows. Section $2$ includes algorithms used for encoding and decoding data into DNA. Section $3$ provides an overview of Graphical User Interface (GUI). Section $4$ describes detailed GUI while Section $5$ has remarks on limitations and assumptions in the software. Section $6$  concludes with challenges in the area of synthetic hard drive and last section provides a link for downloading the software and related material.

\begin{table*}[ht]
\caption{How big is the Big Data?}
\raggedleft
\begin{tabular}{|l|l|l|l|l|}
\hline
Data Unit &  Size & How big it is & On what it can be stored \cite{CompareByte} & Remarks \cite{Howmuchinfo} \\ \hline \hline
Tera Byte (TB) & $1000$ GB  & $200000$ Photos & $1$ TB Hard Disk & $400$ Terabytes: National \\
&             &            &                  &Climactic Data Center (NOAA) database  \\ \hline
Peta Byte (PB)& $1000$ TB  & $3$ years of EOS data & $16$ Backblaze storage pads   & $200$ Petabytes: \\
              &            &  (NASA's Earth Observing System) & racked in two datacenter & All printed material.\\ \hline
Exa Byte (EB)& $1000$ PB   & $2$ Exabytes: Total volume of  & A city Block of $4$  & $5$ Exabytes: All words ever \\
           &             & information generated in $1999$ & storey datacentre & spoken by humans.\\ \hline
Zetta Byte (ZB)& $1000$ EB & $1.9$ zettabytes of information sent & $20$ percent of Manhattan, & $5$ Zetta Byte is equal to \\
&                          & through broadcast technology like T.V and GPS. \cite{zettabyte} & New york & US NSA's Utah Data Center \cite{zettabyte}\\ \hline
Yotta Byte (YB)& $1000$ ZB & $1$ YB is the total Volume of & State of Delware and  & $1.3$ zettabytes is of traffic \\ 
&                          & government data the NSA (National Security Agency) \cite{Yottabyte} & Rhode Island with million Data centre \cite{Yottabyte}& annually over the internet in $2016$ \cite{Yottabyte2016}\\ \hline
\end{tabular}
\label{data}
\end{table*}
\section{Algorithms for Encoding and Decoding Data Files}
While implementing the methods of   \cite{goldman2013towards}, we modified the algorithms little bit so that they are memory efficient. For encoding, algorithm \ref{algo1} generates DNA string from given data file which is further divided into DNA chunks of lengths $117$ using algorithm \ref{algo2}.  For decoding, algorithm \ref{algo3} takes the DNA file containing DNA chunks of length $117$ and produces DNA string which is further decoded to get the original data file using algorithm \ref{algo4}. In order to describe these algorithms we define a  term index info and also give remarks for algorithms \ref{algo3} and \ref{algo4}.
\begin{definition}(Index Info):
 Index info is base 3 string of length 15 which has format (ID: no of chunk : parity of the chunk), where ID has length 2, no of chunks has length  12 and parity of chunk has length 1\cite{goldman2013towards}.  Later on, every chunk is also appended with 'G' or 'C' and prepended with 'A' or 'T' .
\end{definition}
\begin{algorithm}
\caption{Algorithm for generating DNA string}
\begin{algorithmic}
\REQUIRE File size and chunk size
\ENSURE  DNA string for the file 
\STATE$1:$ \IF{ file size $<$ chunk size}
	\STATE Do not divide file into chunks
	\ELSE
	\STATE Divide file into chunks where no of chunks = file size / chunk size + 1
	\ENDIF	
\STATE$2:$ 
\IF {no of chunks $> 1$}
	\STATE Read bytes string from chunk one
	\STATE Convert the string to ascii values list
	\STATE Convert the ascii values list to base 3 string
	\STATE Convert base 3 string to DNA bases and store the last base of the DNA string
		
	\STATE$3:$
	\FOR{chunk number 1 to total number of chunks - 1}
	\STATE Read the bytes string for the chunk
	\STATE Convert the bytes string to ascii values list
	\STATE Convert the ascii value list to base 3 string
	\STATE Convert base 3 string to DNA bases using pervious chunk's last base
	\STATE Concatante this new DNA string with original one
	\STATE Store the last base of the DNA string
 	\ENDFOR
\STATE$4:$  
\STATE Read bytes string for the last chunk
	\STATE Convert the bytes string to ascii values list
	\STATE Convert this ascii value list to base 3 string
	\STATE Convert base 3 string to DNA bases using pervious chunk's last base
	\STATE Concatenate the new DNA string  with original one
\ELSE 
	\IF {no. of chunks = 1}
	\STATE Read entire file and perform conversion steps directly for converting to ASCII then to base 3 then to DNA base (trivial case)
	\ENDIF
\ENDIF
\STATE$5.$ Convert length of the final DNA String obtained to base three and add leading zeros unless length is 20
 \STATE$6.$  Add zeros in between the DNA string and base 3 string obtained in previous step such that total string length is divisible by 25
 \STATE$7.$  Convert the remaining base 3 string to DNA base
\end{algorithmic}
\label{algo1}
\end{algorithm}

\begin{algorithm}
\caption{Algorithm for generating DNA chunks}
\begin{algorithmic}
\REQUIRE DNA string for the file obtained in algorithm \ref{algo1}
\ENSURE  DNA chunks of length 117
\STATE$1:$ \IF{ file size $<$ chunk size}
	\STATE Do not divide file into chunks
	\ELSE
	\STATE Divide file into chunks where no of chunks = file size / chunk size + 1
	\ENDIF	
\STATE$2:$ 
\IF {no of chunks $> 1$}
	\STATE Read DNA string for the chunk one
	\STATE Divide the string into chunks of length 100 and add index info in these chunk 
	\STATE Store last 75 DNA bases of the DNA string read
			
	\STATE$3:$
	\FOR{chunk number 1 to total number of chunks - 1}
	\STATE Read DNA string in the chunk
	\STATE Append the read DNA string to last stored temporary DNA string of 75 bases 
	\STATE Again divide the string into chunks  of length 100, add index info and store its last 75 DNA bases
	\STATE Concatenate the list of chunks to original list of chunks	
	\ENDFOR

	\STATE$4:$  
	\STATE Read last chunk, append the read DNA string to last stored temporary DNA string of 75 bases
	\STATE  Divide the string into chunks of length 100 and add index info in these chunks
	\STATE Concatenate the list of chunks to original list of chunks
\ELSE 
	\IF {no. of chunks = 1}
	\STATE Read entire file and divide the string into chunks of length 100 and add index info in these chunks (trivial case)
	\ENDIF
\ENDIF
\STATE This entire list obtained is stored in .dnac file
\end{algorithmic}
\label{algo2}
\end{algorithm}

\begin{algorithm}
\caption{Algorithm for regenerating DNA string from DNA chunks}
\begin{algorithmic}
\REQUIRE  .dnac file containing DNA chunks of length 117 obtained from algorithm \ref{algo2}
\ENSURE  DNA string for the chunks
\STATE$1:$ \IF{ file size $<$ chunk size}
	\STATE Do not divide file into chunks
	\ELSE
	\STATE Divide file into chunks where no of chunks = file size / chunk size + 1
	\ENDIF	
\STATE$2:$ 
\IF {no of chunks $> 1$}
	\STATE Decode the given chunks read to corrosponding DNA string if possible*
	\WHILE{not decoded}
	\STATE Remove last base from buffer of .dnac file and try decoding again
	\STATE Store this base at the end of prepend string if a bit is removed
	\ENDWHILE
	
	\STATE$3:$
	\FOR{chunk number 1 to total number of chunks - 1}
	\STATE Prepend last stored String to buffer read if prepend string is not null
	
	\WHILE{not decoded}
	\STATE Remove last base from buffer of .dnac file and try decoding again
	\STATE Store this base at the end of prepend string if a bit is removed
	\ENDWHILE
	
	\STATE write the DNA string obtained to original
	\ENDFOR

	\STATE$4:$  
	\STATE Prepend last stored String to buffer read if prepend string is not null, decode it and append it to original DNA string 
\ELSE 
	\IF {no. of chunks = 1}
	\STATE Trivial case so read entire file at once and convert it to DNA string
	\ENDIF
\ENDIF
\end{algorithmic}
\label{algo3}
\end{algorithm}

\begin{remark}(For * in Algorithm~\ref{algo3})
The decoding is always not possible since the format of .dnac file is ['$chunk_1$','$chunk_2$',...,'$chunk_n$']. Now while reading x chunks it may happen that last chunk is not completely read, hence we keep on removing the last byte from the read string unless we get ',' before which entire chunk is decodable.
\end{remark}

\begin{algorithm}
\caption{Algorithm for regenerating original file from DNA chunks}
\begin{algorithmic}
\REQUIRE DNA string obtained from algorithm \ref{algo3}
\ENSURE  Original computer file
\STATE$1:$ \IF{ file size $<$ chunk size}
	\STATE Do not divide file into chunks
	\ELSE
	\STATE Divide file into chunks where no of chunks = file size / chunk size + 1
	\ENDIF	
\STATE$2:$ 
\IF {no of chunks $> 1$}
	\STATE Read DNA string for the chunk 1
	\STATE Convert the DNA string to base 3 string
	\STATE Convert the base 3 string to list of Huffman values if possible**
	\WHILE{not decoded}
	\STATE Remove last base and try decoding again
	\STATE Add removed base to 'prepend string'
	\ENDWHILE
	\STATE Convert the huffman list to corrosponding ascii list
	\STATE Convert ascii list to string of bytes and write to file
	
	\STATE$3:$
	\FOR{chunk number 1 to total number of chunks - 1}
	\STATE Read DNA string for the chunk and prepend 'prepend string' to it if not null
	\STATE Convert the DNA string to list of huffman values if possible
	
	\WHILE{not decoded}
	\STATE Remove last base and try decoding again
	\STATE Add removed base to 'prepend string'(after clearing)
	\ENDWHILE
	
	\STATE Convert huffman list to corrosponding ascii list
	\STATE Convert the ascii list to string of bytes and write to file
	\ENDFOR

	\STATE$4:$  
	\STATE Read last chunk, convert this to base 3 string, to corresponding Huffman list,  to corresponding ascii list, to string of bytes and write this bytes to file
\ELSE 
	\IF {no. of chunks = 1}
	\STATE Trivial case process entire file at once and convert it to base 3 then to huffman list which in turn is converted to ascii list and then to stream of bytes which are then written to a file
	\ENDIF
\ENDIF
\end{algorithmic}
\label{algo4}
\end{algorithm}

\begin{remark}(For ** in Algorithm~\ref{algo4})
The decoding here also is not always possible since huffman values are either of length $5$ or length $6$. So we keep on removing the last byte from the read string and try decoding again and again unless decoded.
\end{remark}
\section{Graphical User Interface (GUI) Overview}
\begin{figure}
\center
 \includegraphics[scale=.20]{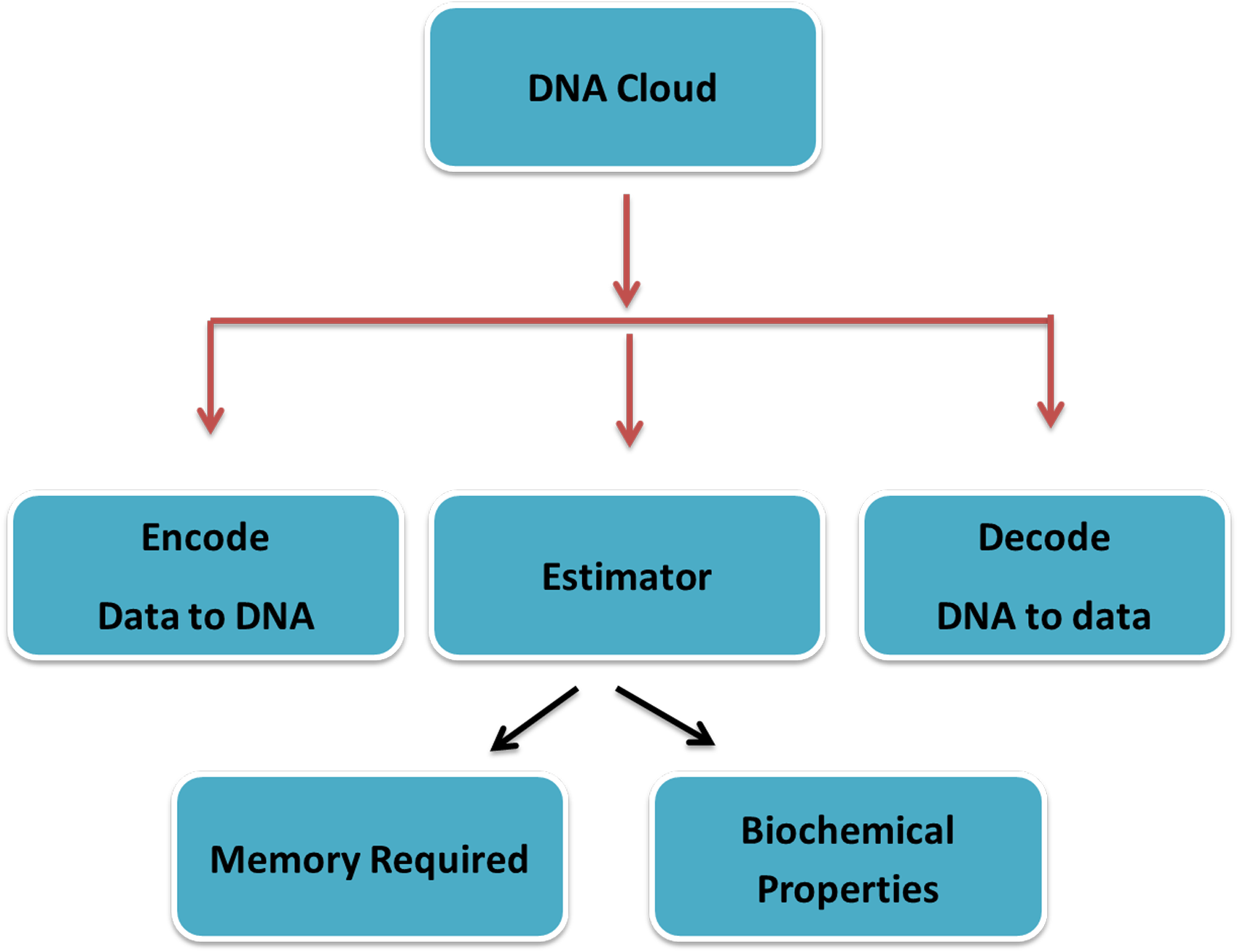}
\caption{Functionality of DNACloud: This flowchart represents the basic function of the DNACloud. As it shows that there are three main modules $1$. Encode,  $2$. Decode and $3$. Estimator. Encode converts the data file of any input and gives DNA sequences as an output. Decode takes DNA sequences as an input and convert it back to original data. Estimator is developed to estimate certain numerical values like memory required for DNA storage that configure your memory of the system and other estimates the biochemical properties of the DNA employed in the wetlab experiments.}
\label{flowchart}
\end{figure}

DNACloud has been primarily developed to facilitate the storage of data on DNA. The software converts any type of data (text, image, audio or video etc.)  into DNA strings and enables it to store on DNA and helps to retrieve the data stored on DNA. The GUI of DNACloud is developed to enable this feature. Along with the encoding and decoding facility, DNACloud provides the user various estimations related to the data storage on DNA as shown in Figure \ref{flowchart}. There are three basic modules of the software as discussed in Sections A, B and C.

\subsection{DNA Encoder (File to DNA)}
To store data on DNA, one has to find ways for encoding the given data into DNA sequence. There are many encoding techniques available to convert the data into DNA sequences by using DNA codes \cite{arita2004writing}. One of the most the efficient source coding technique called Huffman codes is well known for data compression \cite{huffman1952method}. The DNA encoding by Huffman is uniquely decodable. In this software, similar Huffman encoding is implemented  \cite{goldman2013towards}.  For error correction,  the overlapping codes \cite{goldman2013towards} are implemented and data is retrived from DNA with  reduced error rates. The encoding module takes the data file of any format (.text, .png, .jpg, .mp3, .mkv etc.) as an input. The DNA sequence encoded is divided into fixed length of DNA chunks and the part of the DNA chunks were overlapped implementing four fold redundancy for error correction. The original file is converted to Huffman base 3 code (0,1,2) with code length of $5$ which is  transformed to triplet codon to DNA code according to the conversion principle as substituting each trit (triplet) with one of the the three nucleotide different from the preceding one i.e. if G is the preceding, then A or T or C will be placed, this ensures that no homo polymers are generated to reduce the sequencing error. For these code, if any DNA chunk  or base was deleted, then it can be regenerated by reading that overlaped code sequence.  This module saves the encoded file with extension "fileformatextension.dnac". E.g. an image file will be encoded and saved as ".png.dnac".
\subsection{DNA Decoder (DNA to File)}
To retrive the data stored on DNA, data has to be decoded from DNA. The reverse step of encoding is followed for decoding. The data stored in DNA can be retrived by excluding the index bits and converting base $3$ Huffman DNA codes back to original data. This module takes the DNA sequence as input and gives original data stored as output. The output of the sequencer can be used as input for this module. It takes ".dnac" file as input.
\subsection{Storage Estimator}
This module gives various statistics and biochemical properties of DNA for the encoded file. These estimated values are helpful while doing the experiments for storing data in DNA. Estimator has two main sections.\\
\begin{enumerate}
\item \textbf{Memory Required}\\
  User can select the file to be encoded from the system and the following values of the file are estimated. This will help the user to decide how much memory of his system will be occupied for encoding and storing particular data file in DNA. These values are approximated.
\begin{enumerate}
\item File size in bytes
\item Size of DNA string
\item Free Memory Required
\item Amount of DNA Required
\end{enumerate}

\item \textbf{Biochemical Properties and Cost}\\
This will estimate the biochemical properties of the DNA sequence used to store the data. Select ".dnac" file from the system which contains the DNA sequences to estimate the properties. This will take salt concentration (mM) and cost per base as an input. It will estimate the GC content of DNA, melting temperature of the DNA and total cost to store the file in DNA. All the values are approximated. This facilites the user to figure out the budget for the experiments depending on the total amount of DNA.
\end{enumerate}

\begin{table*}[ht]
\caption{Comparision of the file formats encoded by DNACloud. Different file types were encoded and decoded using DNACloud. For cost calculations see \cite{goldman2013towards}}
\raggedleft
\begin{tabular}{|l|l|l|l|l|l|l|l|}
\hline
File\_Type&Limit of File size can be encoded&Encoded File size (Bytes)&Required amount of DNA& Cost of DNA (US \$)& Memory required DNA Chunks& Ref.\\ \hline \hline
Text      & $581130733$ bytes ASCII characters &$15902545$ & $3.4 \times {10}^{-14} $gms & $197191.6$ & $409$MB   & \cite{ribenboim1996recognize} \\ \hline
Audio     & $554$ MB (around $50$ songs)       & $151391203$& $3.3\times {10}^{-13}$ gms& $1877250.9$&  $3896$ MB &  \cite{audiolink}                     \\ \hline
Video     & $581130733.33$ bytes (around $65$ minutes)& $598292824$&$1.31 \times {10}^{-12}$ gms&$7418831.0$& $15400$ MB &   \cite{vediolink}                    \\ \hline
Image (HD)& $5.7$MB (100 HD images) &$23013231$ & $5.051\times {10}^{-14}$ gms & $285364.06 $& $24$MB  &\cite{HDimage}  \\ \hline
\end{tabular}
\label{comparedata}
\end{table*}

\section{DETAILED DESCRIPTION OF GUI}
When the program is executed, a dialouge box is popped up for workspace where one can save his work. All the files generated will be automatically saved in this workspace. You can switch to other workspace. After this dialouge box user details are asked. It includes name, contact number and email address and file you are using as an input. This will save your details and \textbf{Generate Barcode} button will generate a barcode of it which can be used as unique identification by Biotech companies when performing the experiments. It is not mandatory but recommendable to fill the details else the box will remind you again and again. To \textbf{Encode} or \textbf{decode} the file select either of the options from File menu. Software includes options A to G in the menu.
\subsection{Encode (File to DNA) Button}
This option is available under the File menu which will convert any type of data to DNA strings.  User can select the file to be converted into DNA string by clicking on Choose File button. Once the user selects the file to be encoded, the list of information for the encoded file will be displayed as below.
\begin{enumerate}
\item Length of DNA string
\item No of DNA oligonucleotides (chunks)
\item Length of each DNA oligonucleotide
\item File size in bytes
\end{enumerate}
To save the encoded file, user can save the file with specific name on specific location by using Encode your File button. It will generate the file with extension ".dnac" that has DNA string for the file selected. \textbf{RESET} button can clear the selected file then user can select new file. It is like clear button.
\subsection{Decode (DNA to File) Button}
This option is available under the file menu which will retrieve the data stored in DNA. User can enter the DNA string from which data is to be retrieved in the text box against \textbf{Please write DNA string} option. User can also decode the encoded file from the system by option \textbf{Select .dnac file}. Once the file  is selected, click on Decode  option. To save the decoded file, give the name and save the file at specific location. This will generate the original file that was encoded in DNA. \textbf{RESET} button can clear the selected file i.e, It is like a clear button.
\subsection{Storage Estimator}
As mentioned above, it has two estimators. To estimate the memory required user can select the option from \textbf{$File\rightarrow Estimator \rightarrow$Memory Required}. This will take data file to be encoded as input and \textbf{Calculate} button will estimate the values as mentiond above. For second estimator, user can select \textbf{$File\rightarrow Estimator \rightarrow$ }\textbf{Biochemical Properties} option. This will ask ."dnac" file as an input and give the GC content and Melting temperature values and cost for total DNA. \textbf{Save} button will help to save estimated information. 
\subsection{Export Button}
This will help the user to export the file generated to different formats. DNA strings can be exported to file format that can be used as input for the synthesizer. File can be exported to file format that is required by the sequencer. These options are available in File menu. This will generate the feasible output of the DNA strings that is to be used by respective machines. To export the DNA file for the synthesizer to synthesize the DNA, use \textbf{Export DNA synthesizer File} option. To decode the file stored in DNA, use \textbf{Import DNA sequencer file} option to get the DNA sequences to be decoded. These options will be available in the next version of the software. The .dnac file can be exported to PDF and latex file with all software output details in single PDF by using the option \textbf{Export .dnac to PDF} and \textbf{Export to latex} respectively.
\subsection{Clear temp files}
This will clear all your history of the software. It will remove all the temporary files generated by the software.
\subsection{Exit}
Exit will help to quit  the software.
\subsection{User details}
This option is available in preferences menu. For the data security, user has to feed his details then the dialogue box to enter the password appear. Password can be reset with \textbf{Change Password} option in same menu. This helps the user to retrieve his files stored in DNA safely. The barcode generated can be used by biotech companies to tag the DNA on which the particular file is stored.
\section{Comparision of Data Storage on DNA}
The software has limitation of encoding and decoding the file beyond certain file size.  Table \ref{comparedata} compares the file size limit of different file types encoded by the software. At present, the maximum file size of $3486784400$ bytes or $3.4$ GB of DNA strings could be decoded by the software and any file of size $581130733.333$ bytes or $554$ MB can be encoded with DNACloud. 
\section{Conclusion}
Considering the current rate of data explosion, DNA storage becomes an absolutely indispensable data storage medium because of its low maintenance cost, high data density, eco-friendliness and durability. However, the technological advancements are rudimentary, since still the cost for sequencing and synthesizing DNA is pretty high. But since the cost is decreasing every day, we expect that the research in encoding and decoding algorithms can avail common man with this technology within next few years. Thus, DNACloud can be considered as a potential tool to convert data files into DNA and vice versa. We are anticipating to enhance the capability of the software to encode large size data by implementing better encoding and decoding techniques and error correction methods.
\section{SOFTWARE AVAILABILITY}
The software source code, installers for Mac and Windows, user manual, product demo and other related materials can be downloaded from http://www.guptalab.org/dnacloud. 
\section{ACKNOWLEDGEMENT}
We would like to thank Thorsten Weimann and Anand B Pillai whose open source libraries of python-barcode \cite{barcode} and pytxt2pdf \cite{textpdf} respectively are used in the software.
\bibliographystyle{IEEEtran}
\bibliography{dnastoreref}

\begin{thebibliography}{10}
\providecommand{\url}[1]{#1}
\csname url@samestyle\endcsname
\providecommand{\newblock}{\relax}
\providecommand{\bibinfo}[2]{#2}
\providecommand{\BIBentrySTDinterwordspacing}{\spaceskip=0pt\relax}
\providecommand{\BIBentryALTinterwordstretchfactor}{4}
\providecommand{\BIBentryALTinterwordspacing}{\spaceskip=\fontdimen2\font plus
\BIBentryALTinterwordstretchfactor\fontdimen3\font minus
  \fontdimen4\font\relax}
\providecommand{\BIBforeignlanguage}[2]{{%
\expandafter\ifx\csname l@#1\endcsname\relax
\typeout{** WARNING: IEEEtran.bst: No hyphenation pattern has been}%
\typeout{** loaded for the language `#1'. Using the pattern for}%
\typeout{** the default language instead.}%
\else
\language=\csname l@#1\endcsname
\fi
#2}}
\providecommand{\BIBdecl}{\relax}
\BIBdecl

\bibitem{goldman2013towards}
N.~Goldman, P.~Bertone, S.~Chen, C.~Dessimoz, E.~M. LeProust, B.~Sipos, and
  E.~Birney, ``Towards practical, high-capacity, low-maintenance information
  storage in synthesized {{DNA}},'' \emph{Nature}, 2013.

\bibitem{CompareByte}
\BIBentryALTinterwordspacing
G.~Budman. {NSA} might want some backblaze pods. [Online]. Available:
  \url{http://blog.backblaze.com/2009/11/12/nsa-might-want-some-backblaze-pods%
/}
\BIBentrySTDinterwordspacing

\bibitem{reviewpaper}
L.~Dixita and M.~K. Gupta, ``Natural data storage on {DNA}: A review,'' 2013,
  preprint.

\bibitem{davis1996microvenus}
J.~Davis, ``Microvenus,'' \emph{Art Journal}, vol.~55, no.~1, pp. 70--74, 1996.

\bibitem{Genesis}
\BIBentryALTinterwordspacing
E.~Kac. (1999) Genesis-art of {{DNA}}. [Online]. Available:
  \url{http://www.ekac.org/geninfo.html}
\BIBentrySTDinterwordspacing

\bibitem{yachie2007alignment}
N.~Yachie, K.~Sekiyama, J.~Sugahara, Y.~Ohashi, and M.~Tomita,
  ``Alignment-based approach for durable data storage into living organisms,''
  \emph{Biotechnology progress}, vol.~23, no.~2, pp. 501--505, 2007.

\bibitem{portney2008length}
N.~G. Portney, Y.~Wu, L.~K. Quezada, S.~Lonardi, and M.~Ozkan, ``Length-based
  encoding of binary data in {{DNA}},'' \emph{Langmuir}, vol.~24, no.~5, pp.
  1613--1616, 2008.

\bibitem{ailenberg2009improved}
M.~Ailenberg and O.~D. Rotstein, ``An improved huffman coding method for
  archiving text, images, and music characters in {DNA},''
  \emph{Biotechniques}, vol.~47, no.~3, p. 747, 2009.

\bibitem{wong2003organic}
P.~C. Wong, K.-k. Wong, and H.~Foote, ``Organic data memory using the {{DNA}}
  approach,'' \emph{Communications of the ACM}, vol.~46, no.~1, pp. 95--98,
  2003.

\bibitem{arita2004secret}
M.~Arita and Y.~Ohashi, ``Secret signatures inside genomic {{DNA}},''
  \emph{Biotechnology progress}, vol.~20, no.~5, pp. 1605--1607, 2004.

\bibitem{skinner2007biocompatible}
G.~M. Skinner, K.~Visscher, and M.~Mansuripur, ``Biocompatible writing of data
  into {{DNA}},'' \emph{Journal of Bionanoscience}, vol.~1, no.~1, pp. 17--21,
  2007.

\bibitem{church2012next}
G.~M. Church, Y.~Gao, and S.~Kosuri, ``Next-generation digital information
  storage in {{DNA}},'' \emph{Science}, vol. 337, no. 6102, pp. 1628--1628,
  2012.

\bibitem{church2012regenesis}
G.~M. Church and E.~Regis, \emph{Regenesis: how synthetic biology will reinvent
  nature and ourselves}.\hskip 1em plus 0.5em minus 0.4em\relax Basic Books,
  2012.

\bibitem{pmid23514938}
A.~O'~Driscoll and R.~D. Sleator, ``{{S}ynthetic {D}{N}{A}: the next generation
  of big data storage},'' \emph{Bioengineered}, vol.~4, no.~3, pp. 123--125,
  2013, [PubMed
  Central:\href{http://www.ncbi.nlm.nih.gov/pmc/articles/PMC3669150}{PMC366915%
0}] [DOI:\href{http://dx.doi.org/10.4161/bioe.24296}{10.4161/bioe.24296}]
  [PubMed:\href{http://www.ncbi.nlm.nih.gov/pubmed/23514938}{23514938}].

\bibitem{Greengard:2013:NAI:2492007.2492013}
\BIBentryALTinterwordspacing
S.~Greengard, ``A new approach to information storage,'' \emph{Commun. ACM},
  vol.~56, no.~8, pp. 13--15, Aug. 2013. [Online]. Available:
  \url{http://doi.acm.org/10.1145/2492007.2492013}
\BIBentrySTDinterwordspacing

\bibitem{Howmuchinfo}
\BIBentryALTinterwordspacing
K.~Swearingen. How much information. [Online]. Available:
  \url{http://chnm.gmu.edu/digitalhistory/links/pdf/preserving/8_5a.pdf}
\BIBentrySTDinterwordspacing

\bibitem{zettabyte}
\BIBentryALTinterwordspacing
M.~Hilbert. How much information is there in the world? [Online]. Available:
  \url{http://news.usc.edu/#!/article/29360/How-Much-Information-Is-There-in-t%
he-World}
\BIBentrySTDinterwordspacing

\bibitem{Yottabyte}
\BIBentryALTinterwordspacing
R.~Thomchick. {NSA} (national security agency) or {FBI} (federal bureau of
  investigation) will have one yottabyte. [Online]. Available:
  \url{http://www.metaholic-musings.com/2013/03/20/brontobytes/}
\BIBentrySTDinterwordspacing

\bibitem{Yottabyte2016}
\BIBentryALTinterwordspacing
S.~Higginbotham. As data gets bigger, what comes after a yottabyte? [Online].
  Available:
  \url{http://gigaom.com/2012/10/30/as-data-gets-bigger-what-comes-after-a-yot%
tabyte/}
\BIBentrySTDinterwordspacing

\bibitem{arita2004writing}
M.~Arita, ``Writing information into {{DNA}},'' in \emph{Aspects of Molecular
  Computing}.\hskip 1em plus 0.5em minus 0.4em\relax Springer, 2004, pp.
  23--35.

\bibitem{huffman1952method}
D.~A. Huffman, ``A method for the construction of minimum-redundancy codes,''
  \emph{Proceedings of the IRE}, vol.~40, no.~9, pp. 1098--1101, 1952.

\bibitem{ribenboim1996recognize}
P.~Ribenboim, ``How to recognize whether a natural number is a prime,'' in
  \emph{The New Book of Prime Number Records}.\hskip 1em plus 0.5em minus
  0.4em\relax Springer, 1996, pp. 19--178.

\bibitem{audiolink}
\BIBentryALTinterwordspacing
J.~Singh. Vakratunda mahakaya-prathameshwara ganadheeshwara. [Online].
  Available:
  \url{http://music.raag.fm/Bhakti_Sangeet/songs-9797-Shri_Ganesh-Jagjit_Singh}
\BIBentrySTDinterwordspacing

\bibitem{vediolink}
\BIBentryALTinterwordspacing
J.~Rover. National geographic television megastructures 53 ultimate skyscraper
  nyc. [Online]. Available: \url{https://www.youtube.com/watch?v=7lV1SQTqhl0}
\BIBentrySTDinterwordspacing

\bibitem{HDimage}
\BIBentryALTinterwordspacing
Wikipedia. {DNA} structure image. [Online]. Available:
  \url{http://upload.wikimedia.org/wikipedia/commons/thumb/d/d8/Benzopyrene_DN%
A_adduct_1JDG.png/433px-Benzopyrene_DNA_adduct_1JDG.png}
\BIBentrySTDinterwordspacing

\bibitem{barcode}
\BIBentryALTinterwordspacing
T.~Weimann. Code for barcode. [Online]. Available:
  \url{https://bitbucket.org/whitie/python-barcode}
\BIBentrySTDinterwordspacing

\bibitem{textpdf}
\BIBentryALTinterwordspacing
A.~Pillai. Convert text to pdf. [Online]. Available:
  \url{http://code.activestate.com/recipes/189858-python-text-to-pdf-converter%
/}
\BIBentrySTDinterwordspacing

\end{thebibliography}
\end{document}